\documentclass[a4paper,11pt]{article}
\pdfoutput=1 
\usepackage{jinstpub}

\usepackage[T1]{fontenc} 

\usepackage{graphicx}
\RequirePackage{xspace}
\usepackage{hyperref}

\def\epem {\ensuremath{e^+e^-}\xspace}

\newcommand{\kev}{\ensuremath{\mathrm{\,ke\kern -0.1em V}}\xspace}
\newcommand{\mev}{\ensuremath{\mathrm{\,Me\kern -0.1em V}}\xspace}
\newcommand{\mevcc}{\ensuremath{{\mathrm{\,Me\kern -0.1em V\!/}c^2}}\xspace}
\newcommand{\gev}{\ensuremath{\mathrm{\,Ge\kern -0.1em V}}\xspace}
\newcommand{\gevcc}{\ensuremath{{\mathrm{\,Ge\kern -0.1em V\!/}c^2}}\xspace}
\newcommand{\tev}{\ensuremath{\mathrm{\,Te\kern -0.1em V}}\xspace}
\newcommand{\tevcc}{\ensuremath{{\mathrm{\,Te\kern -0.1em V\!/}c^2}}\xspace}
\newcommand{\ev}{\ensuremath{\mathrm{\,e\kern -0.1em V}}\xspace}

\def\m   {\ensuremath{{\rm \,m}}\xspace}

\def\mm   {\ensuremath{{\rm \,mm}}\xspace}
\def\mum  {\ensuremath{{\,\mu\rm m}}\xspace}
\def\nm   {\ensuremath{{\rm \,nm}}\xspace}

\def\cms  {\ensuremath{{\rm \,cm}^{-2} {\rm s}^{-1}}\xspace}

\newcommand{\ener} {\ensuremath{\mathcal{E}}\xspace}

\newcommand{\eT}{\ensuremath{\epsilon T}\xspace}

\newcommand{\be}{\begin{equation}}
\newcommand{\ee}{\end{equation}}
\newcommand{\bc}{\begin{center}}
\newcommand{\ec}{\end{center}}
\newcommand{\bi}{\begin{itemize}}
\newcommand{\ei}{\end{itemize}}
\newcommand{\ben}{\begin{enumerate}}
\newcommand{\een}{\end{enumerate}}

\title{\boldmath A high-luminosity superconducting twin \epem\ linear collider with energy recovery }

\author[a,b]{V.~I.~Telnov}

\affiliation[a]{Budker Institute of Nuclear Physics,\\Novosibirsk, Russia}
\affiliation[b]{Novosibirsk State University,\\Novosibirsk, Russia}

\emailAdd{telnov@inp.nsk.su}

\abstract{Superconducting technology makes it possible to build a high energy \epem linear collider with energy recovery (ERLC) and reusable beams. To avoid parasitic collisions inside the linacs, a twin (dual) LC is proposed. In this article, I consider the principle scheme of the collider and estimate the achievable luminosity, which is limited by collision effects and available power. Such a collider can operate in a duty cycle (DC) and in a continuous (CW) modes, if sufficient power. With current SC Nb technology ($T=1.8$ K, $f_{\rm RF}=1.3$ GHz, used for ILC) and with power $P= 100$ MW, a luminosity $L \sim 0.33 \times10^{36}$ \cms is possible at the Higgs factory with $2E_0=250$ GeV. Using superconductors operating at 4.5 K with high $Q_0$ values, such as Nb$_3$Sn, and $f_{\rm RF}=0.65$ GHz, the luminosity can reach  $L \sim 1.4 \times10^{36}$ \cms at $2E_0=250$ GeV (with P=100 MW) and  $L \sim 0.8 \times10^{36}$ \cms at $2E_0=500$ GeV (with P=150 MW), which is almost two orders of magnitude greater than at the ILC, where the beams are used only once.  This technology requires additional efforts to obtain the required parameters and reliably operation. Such a collider would be the best machine for precision Higgs studies, including the measurement of Higgs self-coupling. }


\arxivnumber{2105.11015}

\begin{document}
\maketitle
\flushbottom

\section{Introduction}
 Linear \epem\ colliders (LC) have  been actively developed since the 1970s as a way to reach higher energies. Their main advantage over storage rings is the absence of synchrotron radiation during acceleration, which makes it possible to achieve much  higher energies. Their main weak point is the one-pass use of beams. At storage rings, the same beams are used many millions of times, whereas in LC  they are sent  to beam dumps after a single collision. This inefficient use of electricity results in a low collision rate and therefore a lower luminosity.

  There were many LC projects in the 1990s (VLEPP, NLC, JLC, CLIC, TESLA, etc.)~\cite{Loew}; since 2004 only two remain: ILC~\cite{ILCTDR,ILC} and CLIC~\cite{CLIC}. The ILC is based on superconducting (SC) Nb technology (in the footsteps of the TESLA), while the CLIC  uses Cu cavities and operates at room temperature. Both colliders work in pulse mode; their beam structures are given in Table 1. The difference is only  in the length of  bunch trains: for the ILC it is 4150 times longer. The luminosities and wall plug powers are very similar. In fact, the use of superconducting technology does not provide significant benefits to the ILC in luminosity. Moreover, the accelerating gradient in the ILC is 2--3 times lower. Although there are some technical advantages - greater spacing between bunches (which is good for detectors), lower peak RF power and lower tolerance requirements.
\begin{table}[!hbtp]
\vspace{3mm}
{
\begin{center}

\caption{Pulse structure of the ILC~\cite{ILC} and CLIC~\cite{CLIC}.}
\vspace{4mm}
\begin{tabular}{| l | l | l |  }  \hline
 & ILC & CLIC  \\ \hline
$2E_0$, GeV & 250 & 250 \\
bunches/train, $n_b$  & 1312 & 354 \\
bunch spacing, ns/m & 554/166 & 0.5/0.15 \\
train length, $\mu$s/km & 727/220 & 0.177/0.053 \\
rep. rate, Hz & 5 & 50  \\
collision rate, kHz & 6.56 & 17.7 \\
power (wall plug), MW & 129 & 225 \\
luminosity, $ 10^{34}\cms$ & 1.35 & 1.37 \\
\hline
\end{tabular}
\end{center}
\vspace{-4mm}
}
\label{Table1}
\end{table}

\begin{figure}[!htb]
\centering
\includegraphics[width=12.5cm]{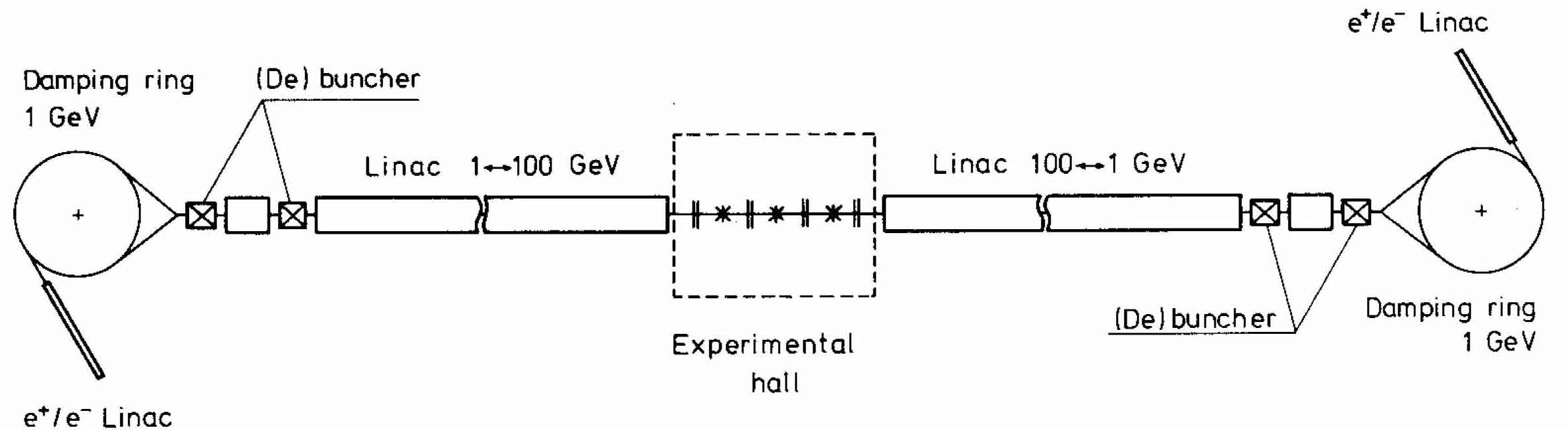}
\caption{Gerke--Steffan's scheme of a linear collider with energy recovery~\cite{Gerke}}
\label{Gerke-scheme}
\end{figure}

The main advantage of the superconducting technology is the feasibility of energy recovery, where the beam, after passing the interaction point (IP), is decelerated in the opposing linac and thus returns its energy to the accelerator. This opportunity was noticed originally and discussed in the very first publications on linear colliders by M.~Tigner~\cite{Tigner}, A.~Skrinsky~\cite{Skrinsky}, U.~Amaldi~\cite{Amaldi}. The LC scheme with energy recovery considered by H.~Gerke and K.~Steffen in 1979~\cite{Gerke} is shown in Fig.~\ref{Gerke-scheme}. This scheme provides not only energy recovery but also multiple use of the same electron and positron beams. One of the problems with multiple use of beams is a large energy spread that appears due to beamstrahlung at the IP. After beam deceleration, the relative energy spread becomes too large for injection to the damping rings. In order to reduce it, authors have foreseen "de(bunchers)" (bunch compressors and decompressors) that change the energy spread and the bunch length while keeping  $\sigma_E \sigma_z$ constant.

  However, the Gerke-Steffen scheme has as few deficiencies:
\bi
\item the quality factor of SC cavities at that time was $Q_0 \sim 2\times 10^9$, which was not enough for the continuous operation mode. Removal of the heat from cryogenic structures requires a lot of energy due to Carnot efficiency; therefore, a duty cycle of 1/30 was adopted.
\item in order to exclude parasitic bunch collisions inside the linac, only one bunch is present at any one moment at each  half linac, which limits the collision rate to of $f=30$ kHz (for a total LC length of 10 km). With a duty cycle of 1/30, the average rate would be a mere 1 kHz.
\item electron and positron bunches cannot be focused by the same final focusing systems (no one had noticed this obstacle), so this scheme could work (which is not obvious) only in one direction of the beams.
\item  the estimated luminosity was $L=3.6\times 10^{31}$ \cms, which is too low to be of interest.
\ei
   Since the 1980s, LC energy recovery schemes have no longer been considered. This is because the collision rate at a single-pass LC is similar to that at an ERL collider (as discussed above), and the luminosity per collision can be much higher at a single-pass LC due to the larger permissible disruption of the beams.

   For many years, linear colliders have been considered as the obvious next large HEP project after the LHC. People expected very rich new physics to emerge in the energy range covered by LCs ($2E_0=$100--3000 GeV). Unfortunately, the LHC has found only the Higgs boson. Physicists are therefore in doubt about linear colliders. It is only absolutely clear that we need an $\epem$ Higgs factory at the energy $2E_0=250$ GeV. But after the discovery of the light Higgs boson,  the FCC-ee and CEPC circular 100 km $\epem$ colliders came into play, promising an order of magnitude higher luminosity at this energy,  followed by the $ 2 \times 100 $ TeV proton collider in the same tunnel.  This is one of reasons (beside the cost) that a decision on the ILC has not yet been made, although it was expected shortly after the publication of the TDR in 2013.

   Below, we revisit the concept of an energy-recovery LC and show that the above problems can be overcome. In addition, significant progress has been made on SC cavities over the past three decades. The quality factor has been increased by more than an order of magnitude. The emphasis will be on the Higgs boson energy, the case of $2E_0=500$ GeV will be considered as well. The result is intriguing and can change the course of the game.

 The outline of the article is the following. I first explain why parasitic collisions should be avoided and suggest a way to solve this problem. In the Section 3 and 4, the collision effects limiting the luminosity are analyzed and  expressions for the luminosities are obtained. Sections 5 and 6 are devoted to the main sources of energy consumptions associated with RF losses in the SC cavities and beam losses to HOM (high order modes),  where the main power is spent on heat removal. Clause 7 estimates the  active running time  with a duty cycle. Section 8 discusses the optimal value of the duty cycles and the number of particles in the bunch. Methods to reduce power consumptions are discussed in Section 9. Section 10 briefly discusses the problem of increasing energy from 250 to 500 GeV. Then, in Section 11, summary tables and figures are presented with the dependencies of the luminosities on the total power for $2E_0=250$ and 500 GeV for several variants of SC cavities.
\section{Superconducting twin linear collider with energy recovery}
   The first question is why it is necessary to exclude parasitic beam collisions inside the linacs. At first glance, the transverse beam sizes in linacs are much larger than at the IP, so  the loss of particles due to such collisions appears to be insignificant. The reason lies in the instability of the beams. If we want to use beams multiple times, the instability criteria are the same as in storage rings and are determined by the vertical tune shift (or the beam-beam parameter)~\cite{Wiedemann}
\be
 \xi_y=\frac{Nr_e\beta_y}{2\pi\gamma\sigma_x\sigma_y} \lesssim 0.1, \;\;\;\;\;\sigma_i= \sqrt{\epsilon_{n,i}\beta_i/\gamma}.
\ee
The ratio of the beam-beam parameters in the linac and at the IP is
\be
\frac{\xi}{\xi^*}=\frac{\sqrt{\beta_y/\beta_x}}{\sqrt{\beta_y^*/\beta_x^*}} \gg 1,
\ee
because in the linac $\beta_x \sim \beta_y$, while at the IP $\beta_x^* \gg \beta_y^*$. Note that this result is independent of the energy at which parasitic collisions occur.

\begin{figure}[!htb]
\centering
\includegraphics[width=15.5cm]{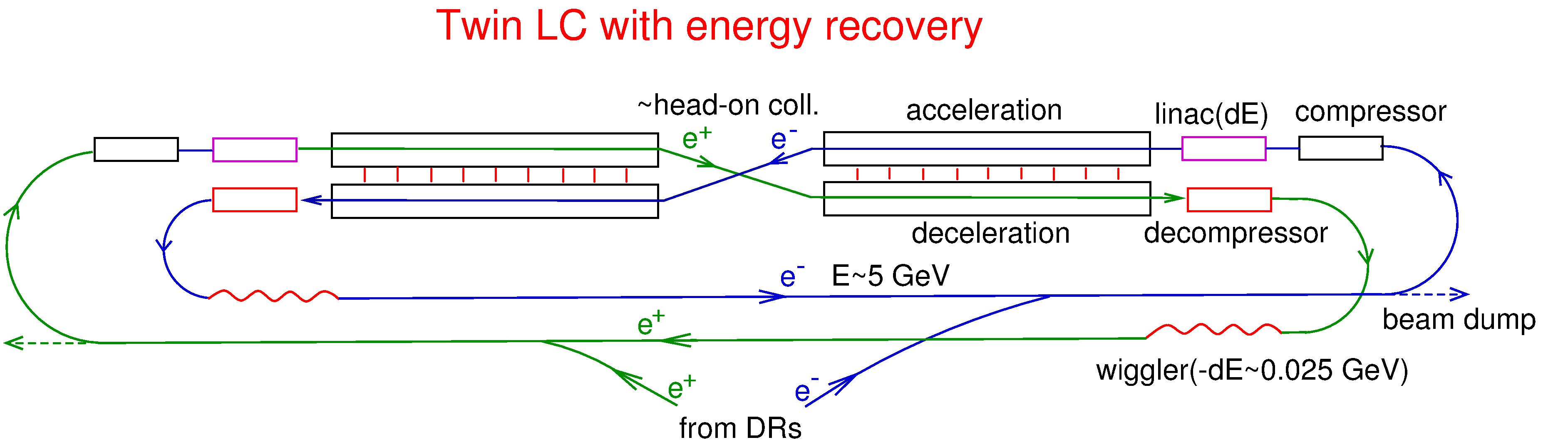}
\caption{The layout of the SC twin linear collider.}
\label{twin-scheme}
\end{figure}

  To solve this problem I propose a twin linear collider in which the beams are accelerated and then decelerated down to $E\approx 5$ GeV  in separate parallel linacs with coupled RF systems, see Fig.~\ref{twin-scheme}. RF power is always divided equally among the linacs. RF energy comes to the beams  both from an external RF source and from the decelerating beam. These can be either two separate SC linacs connected by RF couplers at the ends of multi-cell cavities (9-cell TESLA cavity), or one linac consisting of twin (dual) cavities with  axes for two beams. Such cavities have been designed and tested for XFELs~\cite{cav1,cav2,cav3,cav4}.

  It is assumed that the collider will operate at an energy $2E_0 \approx 250$ GeV (with possible increasing up to 500 GeV)  in a semi-continuous mode with a duty cycle (DC): collisions for about 10 seconds, then a break to cool the cavities.  In one cycle, the beams make about 50 thousand  revolutions. Continuous (CW) operation is more attractive, it is also possible after additional R\&D (discussed below).

    Beams are prepared in damping rings, as usual. In continuous mode, you only need to add the lost particles. In the DC mode, the bunches are prepared anew each cycle. The number of bunches in the ERLC is large, but there is enough time (injection time up to 1-2 s). The required average production rate is an order of magnitude lower than at the ILC.

   During collisions, beams get an additional energy spread that is damped by wigglers installed in the  return pass at the energy $E \approx 5$ GeV. The relative energy loss in wigglers is about $\delta E/E \sim 1/200$. We require that the steady-state  equilibrium energy spread at the IP due to beamstrahlung is better than $\sigma_E/E_0 \sim 0.2\%$, the same as at the ILC and CLIC before the beam collision. Such a spread would be sufficient for beam focusing. The beam parameters are determined not only by the energy spread at the IP, but also by beam lifetime which is caused by single beamstrahlung. We will consider both effects.

When the beam is decelerated down to 5 GeV, its relative energy spread increases by $E_0/E\sim 25$ times  to $\sigma_E/E\sim 5\%$. To make it acceptable for travel without losses in the arcs, its energy spread is reduced by 10--15 times with the help of the bunch (de)compressor; then, the relative energy spread in the arcs will be less than 0.5\%. The beam lifetime will be determined by the tails in beamstrahlung radiation. This loss should not exceed about 1\% after 10000 revolutions. The IP energy spread, the beam instability and beam particle losses (due to single beamstrahlung) determine the IP beam parameters, and hence the luminosity.

  An important question is the injection and extraction of the beams. When the collider is full, the distance between bunches is small, optimally equal to the linac's RF-wavelength ($\lambda_{RF}=23$ cm for TESLA/ILC 1.3 GHz cavities); they are accelerated and decelerated due to the exchange of energy between the beams. External RF power is required only for energy stabilization  and compensation for radiation and high order mode (HOM) losses. During the injection/extraction of the beams, normal energy exchange does not occur until the bunches fill the entire orbit, so the external RF system must work at full power. However, at the ILC, the power of the RF system is only sufficient to accelerate beams with a bunch distance of 100--150 m. In our case, with energy recovery, we need a much shorter inter bunch distance (and smaller bunch charges). This is so necessary because the energy lost by a bunch into high-order modes (HOM) is proportional to $ N^2 $ ($ N $ is the number of particles in the bunch).
    To solve this problem, one must first inject bunches (better short trains) with a large interval and then (at subsequent revolutions) add new trains between the trains already circulating. The distance of 23 cm between the bunches is too small for working with individual bunches, while the use of trains with gaps of 1--2 m makes it possible to manipulate trains using impulse deflectors.  Removal of beams is done in reverse order.
\section{Collision effects limiting the luminosity}
\subsection*{Energy spread in beam collisions}
   During the beam collisions, the particles emit synchrotron radiation (beamstrahlung), which contributes to the energy spread of the beam. The increase of the beam energy spread in a single collision ($n_{\gamma}<1$)~\cite{Yokoya1,Noble,Chen-Yokoya}
\be
\Delta\sigma_E^2=n_{\gamma}\langle \epsilon_{\gamma}^2\rangle=\frac{\langle \epsilon_{\gamma}^2\rangle}{\langle \epsilon_{\gamma}\rangle^2}\frac{\left(n_{\gamma}\langle \epsilon_{\gamma}\rangle \right)^2}{n_{\gamma}}\approx \frac{5.5\,(\Delta E)^2}{n_{\gamma}},
 \ee
\be
\frac{\Delta E}{E_0} \approx \frac{0.84r_e^3N^2\gamma}{\sigma_z\sigma_x^2}, \; n_{\gamma}\approx 2.16\frac{\alpha r_e N}{\sigma_x},
\ee
where $\Delta E$ is the average energy loss, $n_{\gamma}$ is the average number of photons per collision ($n_{\gamma} \ll 1$ under our conditions), $\alpha=e^2/\hbar c\approx 1/137$, $r_e=e^2/mc^2$.  Here, we  neglect the energy spread due to the inhomogeneity of the Gaussian beam ($\sigma_E\approx 0.54\Delta E$), which is much smaller. As a result, we get
\be
 \frac{\Delta \sigma_E^2}{E_0^2}\approx 1.8\frac{N^3r_e^5\gamma^2}{\alpha \sigma_x^3\sigma_z^2}.
\label{deltae}
\ee
   After the collision the bunch decelerates and then stretches during the bunch decompression where its $\Delta\sigma_E$ and $\sigma_E$ decrease proportionally. Due to  SR radiation in damping wigglers at the energy $E \sim 5$ GeV, where particles lose the energy $\delta E$ in one pass, an equilibrium energy spread is achieved~\cite{Wiedemann}:
\be
      \frac{\Delta \sigma_E^2}{\sigma_E^2}=2\frac{\delta E}{E}.
\label{eqv}
\ee
Substitution of (\ref{deltae}) into (\ref{eqv}) gives the  equilibrium energy spread at the IP
\be
\left(\frac{\sigma_E}{E_0}\right)^2\approx 0.9\frac{N^3r_e^5\gamma^2}{\alpha \sigma_x^3\sigma_z^2 (\delta E/E)}.
\ee
For the desired damping rate and energy spread, we obtain the requirements for the beam parameters at the IP
\be
\frac{N^3}{\sigma_x^3\sigma_z^2}< \frac{8\times 10^{-3}}{r_e^5\gamma^2}\left(\frac{\sigma_E}{E_0}\right)^2\frac{\delta E}{E}.
\label{rest1}
\ee
\subsection*{Beam instability}
Our linear collider behaves as a cyclic storage ring, so there is a second limitation on the beam parameters at the IP, due to the tune shift. For flat beams and head-on collisions, it is
\be
\xi = \xi_y=\frac{Nr_e\sigma_z}{2\pi\gamma\sigma_x\sigma_y} \lesssim 0.1 \;\;\; ({\rm for}\;\; \beta_y\approx\sigma_z).
\label{rest2}
\ee
At the ILC, the beams collide at a crossing angle of $\theta_c \approx 15$ mrad, which makes it easier to remove highly disrupted beams. In the ILC case, you should use the crab-crossing scheme (tilt of the bunches by $\theta_c/2$) to preserve the luminosity. In the case of considered cyclic LC, beam disruption is small and beams can be removed through the aperture of the opposing final quadrupole; therefore we assume nearly head-on collisions with a small crossing angle to facilitate the separation of the beams.

   In this article we do not consider  collisions at a large crossing angle ("crab-waist" scheme) because it would provide no benefit when beamstrahlung is important or the beams are short. Some gain is possible at low energies (at Z-boson); this will be discussed in subsequent articles.
\subsection*{Beam lifetime}
  The beam lifetime at high-energy \epem\ storage rings is determined by the emission of high-energy beamstrahlung photons~\cite{Telnov-restr}. An electron (positron) is lost when its energy loss is greater than $\eta E_0$, where $\eta$ is the energy acceptance. In our case, the bunches are decelerated by a factor of 125/5=25 and then expanded by a factor of $\sim 15$;  therefore, the energy acceptance in the 5 GeV arc should be approximately 25/15=1.67 greater than the maximum acceptable relative energy loss at the IP. If we take the energy acceptance in arcs at 5 GeV  equal to 3\%, then at the IP it should be $\eta=0.03/1.67=0.018$ for $E_0=125$ GeV and 0.018(125/E) for other energies.

 The formulas for calculating the beam lifetime are given in the Ref.~\cite{Telnov-issues}. For the lifetime of the beam to correspond to $n_{\rm col}$ collisions in the collider with energy acceptance $\eta$, it is necessary to have
 \begin{equation}
\frac{N}{\sigma_x \sigma_z} < \frac{3.6\times 10^{-3}\eta}{\gamma {r_e}^2 \ln{(7\times 10^{-7} \eta \sigma_z n_{\rm col}/\gamma r_e)}}
\label{life1}
\end{equation}
 Below, we will show that in the case of operation with a duty cycle, the duration of the active phase can be around 10 s. During this time, each bunch collides about 50 thousand times. To have a beam loss at the level of <3$\% $, the beam lifetime must correspond to $ n_{\rm col} \sim 1.5 \times 10^6 $ (the lifetime of about 5 minutes). Since the dependance on $ n_{\rm col}$ is logarithmic, suppose for further consideration $ n_{\rm col}=1.5\times 10^6$, it will be good also for continuous (CW) operation. As explained above, we assume $\eta =0.018 (125/E)=4400/\gamma$. As for the value of the $\sigma_z$ in the logarithm, the range of acceptable bunch lengths is rather narrow, 0.3--2 mm,  we put to the logarithm $\sigma_z=0.5$ mm. With these assumptions, we obtained the third constraint on beam parameters associated with the beam lifetime due to single beamstrahlung
 \be
 \frac{N}{\sigma_x \sigma_z} < \frac{7.9}{\gamma^2 r_e^2 \Lambda},\;\;\;  \Lambda \approx \ln{\frac{120}{E_0/125}},\;\;\;E_0 \;{\rm in\;GeV}.
\label{rest3}
 \ee
The inaccuracy due to fixed values of the quantities in the logarithm does not exceed 20\%.
\subsection*{Bunch length}
  Below we will consider limitations on luminosity due to the three effects described above. There is an optimal bunch length for each beam energy. For low energies it is shorter, for higher energy it is longer, but the preferred range of bunch lengths for the collider based on 1.3 GHz SC cavities is $\sigma_z=$(0.2--0.3)--(1--2) mm. This will be the fourth constraint which also significantly affects the optimization of the parameters
  \be
  \sigma_{z,{\rm mim}} <\sigma_z <\sigma_{z,{\rm max}}.
  \label{rest4}
  \ee
 \section{Luminosity restrictions due to collision effects}
   For four unknown beam parameters at the IP: $N, \sigma_x, \sigma_y, \sigma_z$ we have four restrictions, (\ref{rest1}), (\ref{rest2}), (\ref{rest3}), (\ref{rest4}) and one relationship: $\sigma_y \approx \sqrt{\epsilon_{ny}\sigma_z/\gamma}$.  Collision effects for flat beams depend  on the combination $N/\sigma_x$; therefore, $N$ as a free parameter, its optimum value is discussed in the following sections.

Now we have to understand the complex problem of finding the luminosity under four constraints on the parameters. Let's enumerate and give short names to these constraints:
\ben
\item (BI) Beam Instability (\ref{rest1}).
\item (MBS) Multiple Beamstrahlung ($\sigma_E$ at the IP) (\ref{rest2}).
\item (SBS) Single Beamstrahlung (beam life time) (\ref{rest3}).
\item (SZ) Bunch length, a) $\sigma_z> \sigma_{z, \rm min}$, b) $\sigma_z < \sigma_{z, \rm max}$ (\ref{rest4}).
\een
With increasing energy, the luminosity is limited by the following combinations of constraints: BI-SZ(a) $\Rightarrow$ BI-MBS $\Rightarrow$ BI-SBS $\Rightarrow$ SBS-SZ(b). The combination MBS-SBS is omitted because MBS and SBS  restrict almost the same combinations: $N/(\sigma_x\sigma_z)$ and $N/(\sigma_x\sigma_z^{3/2})$;  MBS-SZ(b) is omitted because SBS-SZ(b) is stronger at high energies.

The results for these four cases are the following.
\subsection*{BI-SZ}
\be
\sigma_x=\frac{Nr_e\sigma_{z, \rm min}^{1/2}}{2\pi\gamma^{1/2}\epsilon_{ny}^{1/2} \xi}, \;\;\; \sigma_y=\left(\frac{\sigma_{z, \rm min} \epsilon_{ny}}{\gamma}\right)^{1/2}, \;\;\;\; \sigma_z=\sigma_{z, \rm min}.
\ee
\be
L \approx \frac{N^2f}{4\pi\sigma_x\sigma_y}=\frac{Nf\xi\gamma }{2 r_e \sigma_{z, \rm min}}.
\ee
\subsection*{BI-MBS}
\be
\sigma_z\approx 19.2\frac{\xi^{6/7}\epsilon_{ny}^{3/7}r_e^{4/7}\gamma}{(\sigma_E/E_0)^{4/7}(\delta E/E)^{2/7}},
\label{sigz}
\ee
\be
\sigma_x \approx 0.7\frac{Nr_e^{9/7}}{\xi^{4/7}\epsilon_{ny}^{2/7}(\sigma_E/E_0)^{2/7}(\delta E/E)^{1/7}}, \;\;\; \;\;\; \sigma_y= 4.4\frac{\xi^{3/7}\epsilon_{ny}^{5/7}r_e^{2/7}}{(\sigma_E/E_0)^{2/7}(\delta E/E)^{1/7}}.
\ee
\be
 L \approx \frac{N^2f}{4\pi\sigma_x\sigma_y}=2.6\times 10^{-2}\frac{Nf\xi^{1/7}}{\epsilon_{ny}^{3/7}r_e^{11/7}}\left(\frac{\sigma_E}{E_0}\right)^{4/7}\left(\frac{\delta E}{E}\right)^{2/7}.
 \label{lum0}
\ee
For $\sigma_E/E_0=2 \times 10^{-3}$, $\delta E/E=0.5\times 10^{-2}$, $\xi=0.1$, $\epsilon_{ny}=3\cdot 10^{-8}$ m (as at the ILC), we have
\be
\sigma_x\approx 0.9\left(\frac{N}{10^{9}}\right)\;\mu{\rm{m}}, \;\;\;\sigma_z\approx 0.3\frac{E[{\rm GeV}]}{125}\; {\rm{mm}}, \;\;\; \sigma_y=6.1\, {\rm nm}.
\label{sigxyz}
\ee
\be
L\approx 4.35 \times 10^{35}\frac{(N/10^{9})}{d[{\rm{m}}]}\approx 9\times 10^{36} I[{\rm{A}}]\; \cms,
\label{lum}
\ee
where $d = c/f$ is the distance between the bunches. Please note, this luminosity is for  continuous operation (100\% duty cycle).
For example (the choice of $N$ and $d$ will be discussed later),
\be
N=10^{9}, d=\lambda_{RF}=0.23\, {\rm m}\; (I=0.21 {\rm A}) \Rightarrow L \approx 1.9\times 10^{36}\, \cms.
\label{lum2}
\ee
In this example, the power radiated in damping wigglers by both beams is $P_{\rm SR}=$ 10.4 MW.
\subsection*{BI-SBS}
\be
\sigma_z=0.86 \Lambda^{2/3}\epsilon_{ny}^{1/3}(\xi r_e)^{2/3}\gamma^{5/3},    \;\;\;\;\;\Lambda= \ln{\frac{120}{E_0/125}},
\label{bisbs-z}
\ee
\be
\sigma_x=\frac{0.15 \Lambda^{1/3} Nr_e^{4/3}\gamma^{1/3}}{\epsilon_{ny}^{1/3}\xi^{2/3}}, \;\;\;\;\sigma_y=0.93 \Lambda^{1/3} \epsilon_{ny}^{2/3} (\xi r_e \gamma) ^{1/3},
\label{bisbs-xy}
\ee
\be
L=\frac{0.58 Nf \xi^{1/3}}{\epsilon_{ny}^{1/3}r_e^{5/3}\gamma^{2/3}\Lambda^{2/3}}.
\label{bisbs-l}
\ee
For $\xi=0.1$ and $\epsilon_{ny}=3\cdot 10^{-8}$ m
\be
L\approx 4.15 \times 10^{35}\frac{(N/10^{9})}{d[{\rm{m}}]} \left(\frac{\ln{120}}{\ln{(120(125/E_0[\gev]))}}\right)^{2/3}\left(\frac{125}{E_0[\gev]}\right)^{2/3}.
\ee
\subsection*{SBS-SZ}
\be
\sigma_x=\frac{0.125N \Lambda \gamma^2 r_e^2}{\sigma_{z, \rm max}}, \;\;\;\;\sigma_y= (\sigma_{z, \rm max} \epsilon_{ny} /\gamma)^{1/2},\;\;\;\;\sigma_z=\sigma_{z, \rm max},
\ee
\be
L=\frac{0.63Nf\sigma_z^{1/2}}{\Lambda \epsilon_{ny}^{1/2}\gamma^{3/2} r_e^2}.
\ee
\subsection*{Summary of collision effects}
\begin{figure}[!htb]
\centering
\includegraphics[width=14.cm, height=11.5cm]{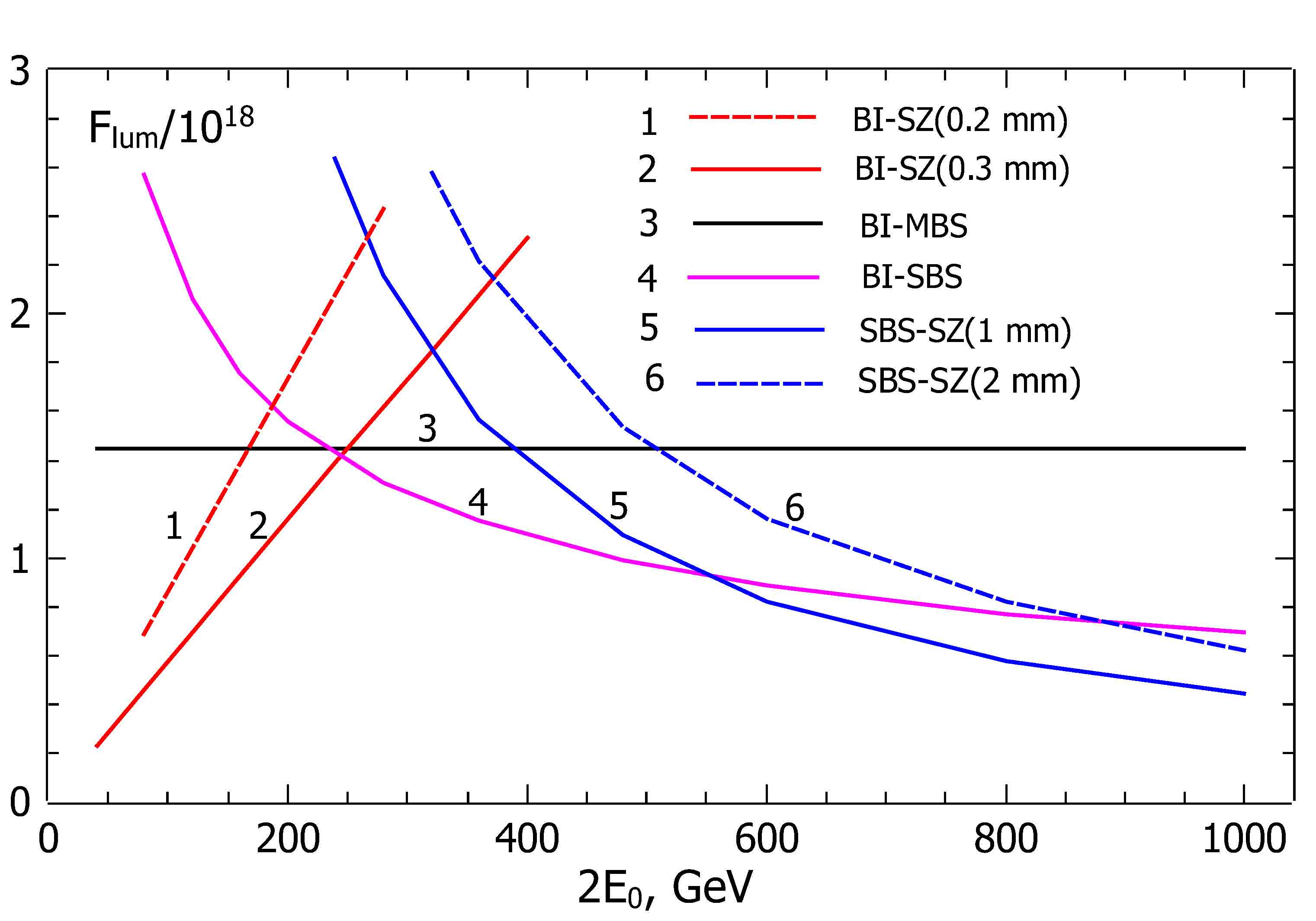}
\caption{Luminosity restrictions due to various collision effects (see the text). The maximum value of $L=Nf\times F_{\rm lum}$, where the limit is given by lowest $F_{\rm lum}$ at each energy.}
\label{lum1234}
\end{figure}

    We have considered various limitations on the luminosity due to collision effects.  The maximum luminosity can be written as
\be
L=(Nf)\,  F(E) \times DC,   \;\;\;\ f=c/d,\;\;\;\;F(E)= \min{F_i}(E),
\ee
where $F_i(E)$ are shown in Fig.~\ref{lum1234}. One can see which effect is most important for a given energy. For $d=\lambda_{\rm RF}= 23$ cm
 \be
L=1.3\times 10^{18} \left(\frac{N}{10^9}\right) F(E) \times DC.
\label{Lopt}
\ee

For $2E_0 = $250--500 GeV the luminosity is limited by BI-SBS (beam instability and single beamstrahlung). Using Eqs.  (\ref{bisbs-z}), (\ref{bisbs-xy}), (\ref{bisbs-l}) we get for $d=23$ cm the following luminosity and beam sizes for $E_0=$ 250 and 500 GeV in continuous (CW) mode:
\be
2E_0\!=\!250:L\!=\!1.81\times 10^{36}(N/10^9),\sigma_x\!=\!0.94(N/10^9)\mum, \sigma_y\!=\!6.2\nm, \sigma_z\!=\!0.31\mm.
\ee
\be
2E_0\!=\!500:L\!=\!1.27\times 10^{36}(N/10^9),\sigma_x\!=\!1.12(N/10^9)\mum, \sigma_y\!=\!7.4\nm, \sigma_z\!=\!0.89\mm.
\ee
For $d=23$ cm the average current $I=0.21(N/10^9)$ A. The optimum value of $N$ and $DC$ is discussed in the Section~\ref{choice}. It will be shown below that the optimum value of $N/10^9\sim$ 0.5--2 for various types of SC cavities.
\section{RF losses in cavities} \label{RF}
    One of the main problems of SC linear accelerators operating in a continuous mode is heat removal from the low-temperature SC cavities. Energy dissipation in one (multi-cell) cavity
\be
P_{\rm RF,dis}=\frac{V_{\rm acc}^2}{(R/Q) Q_0},
\ee
where $V_{\rm acc}$ is the operating voltage and $Q_0$ is the cavity quality factor. The 1.3 GHz TESLA--ILC  cavity has  $R/Q=1036$ Ohm and the length $L_c=1.04$ m.  For the accelerating gradient $G=20$ MeV/m and $Q_0 =3\times 10^{10}$, the thermal power $P_{\rm diss}=13.5$ W/m, or 680 W/GeV.  Also, some heat will add the absorption in SC cavities of the High Order Mode (HOM) losses,  we will take them into account separately in the next section.  Such a continuous-mode SC linac is currently being developed  for the XFEL LCLS-II  at SLAC~\cite{LCLSII,LCLSIIa}, they plan $ P _ {\rm RF,dis} \sim 1 $ kW/GeV.

The overall heat transfer efficiency from temperature $T_2 \approx 1.8$ K to room temperature $T_1\sim 300$ K is $\eta=\varepsilon T_2/(T_1-T_2) \approx 0.18\times 1.8/300 = 1/900$~\cite{eff}. The required refrigeration power for the twin 250 GeV collider is
\be
P_{\rm RF-refr, 2K}\approx 2 \times 250\, {\rm GeV} \times 900 \times 0.68\, {\rm kW/GeV} =306 \, {\rm MW}.
\ee
It is only from RF losses. As we will see below, under optimal conditions, approximately similar power is required for removal of HOM losses.

   In recent years, great progress has been made both in increasing the maximum accelerating voltage and in increasing the quality factor $Q_0$. In the ILC project, it is assumed $Q_0=10^{10}$ and $G=31.5$ MeV/m. For continuous operation, it is advantageous to work at $G \approx 20$ MeV/m, where $Q_0\sim 3\times 10^{10}$ is within reach now. Moreover, N-doping and other surface treatment technologies have already resulted in $Q_0 \sim 5\times 10^{10}$ at $T=2\,$K and $Q_0 \sim (3-4)\times 10^{11}$ at $T<1.5 $ K~\cite{Gonnella,Posen}. According to a leading expert~\cite{Padamsee}, one can hope for a reliable $Q_0=8\times 10^{10}$ at $T=1.8$ K. Currently, we can take $Q_0 \sim 3\times 10^{10}$ and work with duty cycle (DC) in order to keep  the total power consumption of the collider below 150 MW, as at the ILC.

   A even more promising way to reduce the cooling power is to use superconductors with a higher operating temperature, such as $ \rm Nb_3Sn$. Thus, at $T_2 = 4.5$ K the technical efficiency of heat removal is about 30\%~\cite{eff}, and the total one (with Carnot) about 1/220, that is about 4 times higher than that at 1.8 K. This question is discussed also in the Section~\ref{choice}.
\section{High-order mode losses (HOM)} \label{HOM}
    When particles are accelerated in linear accelerators, they take energy $\Delta E = e\ener_0\Delta z$ from the cavity due to the destructive interference of the RF field in the cavity  $\ener_0$ and the wave $\ener_r$ radiated by the bunch into the cavity. When the particles are decelerated ($\Delta E = -e\ener_0\Delta z$), they return their energy back to the cavity  due to constructive interference between the RF field  and the radiated field. However, such a picture with an ideal energy exchange is valid only for the fundamental cavity mode.   High radiation modes (longitudinal wake fields $\propto$ bunch charge) lead to energy losses during both acceleration and deceleration. For this reason, the energy recovery efficiency is not 100\%.

   When the beam passes through a single diaphragm with a radius of $r_a$,  the energy loss can be easily estimated as the energy of the beam field at $r>r_a$. However, for a long linear accelerator with multiple apertures, the picture is more complex. In this case, according to R.~Palmer~\cite{Palmer}, the energy loss by one electron per unit length
\be
-\frac{dE}{dz}\approx \frac{2e^2N}{r_a^2}.
\label{palmer-1}
\ee
It is noteworthy than the energy losses do not depend on the distance between the diaphragms and on the bunch length. This simple formula is supported by detailed numerical calculations. There is a dependence on the bunch length, but very weak.  For TESLA--ILC accelerating structures ($r_a=3.5$ cm), a numerical calculation~\cite{Novokhatsky} gives the  energy losses in the wakefield for $\sigma_z=400$ $\mu$m (the energy loss in the cryomodule is divided by the active length of accelerating cavities)
\be
-\frac{dE}{dz}\approx 2.2 \left(\frac{N}{10^{9}}\right){\rm \frac{keV}{m}}.
\ee
For bunch lengths $\sigma_z=$ 0.25--1 mm, the coefficient varies within the range 2.42--1.76. The formula (\ref{palmer-1}) gives 2.35.
So, for TESLA cavities and $N=10^{9}$, these HOM losses are  $\sim$0.01\% of the accelerating gradient $G\sim$ 20--30 MeV/m.

   For $2E=250$ GeV, $G=20$ MeV/m,  the active collider length is $L=12.5$ km. The total power of HOM energy losses (twin collider, both beams)
\be
P_{\rm HOM}=\frac{2.65}{d[{\rm m}]}\left(\frac{N}{10^{9}}\right)^2\;{\rm MW}.
\ee
For $N=10^{9}$ and $d=0.23$ m $P_{\rm HOM}=11.5$ MW, which is close to the power of synchrotron radiation in damping wigglers, equal to 10.4 MW. Keep in mind that these numbers are for continuous operation.

What happens with the HOM power generated by the beam inside the linear accelerator? For $N=10^9$, this power is about 35 times greater than the RF power dissipation in cavities. Fortunately, most of this energy can be extracted from the SC cavities in  two ways:  a) using HOM couplers which dissipate the energy at room temperature; b) with the help of special HOM absorbers located between the cavities. The latter are maintained at an intermediate temperature around 80~K where refrigeration systems operate at much higher efficiencies.  However, some small part of the HOM energy is dissipated in the walls of SC cavities at 1.8 K. The ratio of the total HOM power to the RF dissipation power
\be
   \frac{P_{\,\rm HOM}}{P_{\rm RF,dis}} \propto \frac{N^2 Q_0}{E_{\rm acc}^2 d}.
\ee
Comparing ILC ($N=2\times 10^{10}, Q_0=10^{10}, d=100 \m$) and ERLC ($N \approx 10^{9}, Q_0=3 \times 10^{10}, d=0.23 \m$) we see that this ratio is  7.9 times higher at the ERLC.  At the ILC $P_{\, \rm HOM}/P_{\rm RF,dis} \approx 4.3$ and about 7\% of HOM power is dissipated at 2 K, so HOM losses adds about 30\%.  For the ERLC, in order for the HOM absorption in the SC cavities to be less than 50\% of the dissipated RF power, it is necessary that the fraction of HOM  power absorbed in the SC resonators does not exceed 1.5\%. High frequency HOMs are not a problem, they propagate like photons, the reflection coefficient from the walls of SC cavities is very high, $1-r<10^{-7}$, so that freely flying photons are very efficiently absorbed by the HOM absorbers located between cavities.  The main difficulty present some trapped low frequency HOM modes with low amplitude at the ends of the cavity where they are damped by HOM couplers.

Here I want to make an important point. Above, we assumed that the HOM loss from the bunch train is the sum of the losses of each bunch separately. However, this is obviously not the case for sufficiently low HOM frequencies. The bunches run equidistantly and the HOM fields emitted into cavities are added with different phases. If the phases were random, then the radiation powers would add up, but in our case there is a strict periodicity. If the HOM frequency is not in resonance with the RF frequency, then the amplitude of the resulting HOM wave from many bunches can only be several times higher than the amplitude from one bunch, which is proportional to $N$  and it does not depend on the repetition rate of the bunches. This means that bunches not only emit, but also take HOM energy from the cavity. For high frequencies, the phase is lost after many reflections due to geometry imperfections, and then the addition of powers is justified, but for low frequencies, which include the captured modes, this is not true. In this case, the HOM power at low frequencies at the ERLC is lower than in the ILC as $N^2$, that is $~(2\cdot10^{10}/10^9)^2 = 400$ times!  This issue is very  important and requires further detailed study. Perhaps we significantly overestimate the HOM power from the bunch train.  In the ILC,  7\% of HOM energy is absorbed in cavities, which is not critical for the ILC.  Proceeding from the above considerations, we take for the ERLC the fraction of HOM losses in SC cavities lower, equal to 1 \%.

This HOM problem in high-current ERL linacs is well-known. The HOM power can be reduced by increasing the iris radius (be decreasing the RF frequency) since  $P_{\rm HOM}\propto 1/r_a^2$ or by decreasing the number of particles in the bunch: the HOM power is proportional to $N^2$, while $L\propto N.$

Let us find the refrigeration power needed for removal of HOM losses assuming that all HOM energy is dissipated in HOM absorbers at a liquid nitrogen temperature of 77 K.  Assuming that refrigeration efficiency is $\eta = 0.3T_2/(T_1-T_2) = 0.3 \cdot 77/(300-77) \sim 0.1$ we find that the refrigeration power for removal of HOM losses is ($d=23$ cm)
\be
P_{\,\rm HOM,refr,77K} = P_{\, \rm HOM}/\eta =10\,P_{\,\rm HOM}\approx (11.5/0.1)(N/10^9)^2 =115(N/10^9)^2\;{\rm MW}.
\ee
One should add to this number the electric power needed for compensation of beam energy losses. Assuming 50\% efficiency it is about $2P_{\, \rm HOM}$.

Less definite are HOM losses in the walls of the SC cavities at 2 K.  Taking the heat transfer efficiency from 1.8 K to 300 K equal 1/900, as in the previous section, and the fraction of HOM losses in cavities equal to 1\% (as discussed above) we find the refrigeration power
\be
P_{\,\rm HOM,refr,2K} = P_{\,\rm HOM} \times 900 \times 0.01 \approx 9\,P_{\,\rm HOM}  \approx 103(N/10^9)^2\;{\rm MW}.
\ee
The total electric power consumption from the wall plug (w.p.) due to HOM losses for $2E_0=250$ GeV collider in the continuous mode of operation
\be
P_{\,\rm HOM, w.p.} \approx 21\, P_{\,\rm HOM} \approx 240 (N/10^9)^2\;{\rm MW}.
\ee
\section{Duration of pulse in DC mode.}  If the refrigeration power needed for continuous operation is too high one can work with a duty cycle $DC \sim$ 0.1--1 (see optimization below).  Duration of continuous operation
is determined by the heat capacity of the liquid He that surrounds the cavity and can be estimated as
\be
\Delta t= \frac{c_p m \Delta T}{P_{diss}} \sim 20 \,{\rm s},
\ee
where $c_p(\rm He)= 2.8$ J/(g$\cdot$K) at T=1.8 K, $m$ is the mass of liquid He per one TESLA cavity (we take 0.02 ${\rm m}^3$ or 2.9 kg), $P_{\rm diss}\sim 20$ W, $\Delta T \sim 0.05$ K. So, we can safely choose the work duration $\Delta t =10$ s.

 More promising for ERLC are superconductors working at $T \approx 4.5$ (temperature of boiling He). The product $\rho c_p$ for this temperature is 1.6 times larger, beside $\Delta T$ can be taken larger by a factor of 3, then  $\Delta t =20 \times 1.6 \times 3 = 96$ s. In this case, the active cycle time  can last about 1 minute.

 Duration of the break is described by the duty cycle $DC$, which depends of the available power, the optimization is given in the next section.
 \section{Optimum values of $N$, distance between bunches $d$  and duty cycle $DC$ } \label{choice}
        Let us find the optimum number of particles in one bunch when the luminosity is maximum for a given power consumption. There are three main energy consumers (numbers correspond to $2E_0=250$ GeV)
        \bi
        \item Radiation in the damping wigglers. $P_{\rm SR}/\varepsilon=20.8(N/10^{9}) \times DC$ MW at $\varepsilon=50\, \%$. This assumes that new beams are generated for each cycle.
        \item Electric power for cooling of the RF losses in cavities (at 2 K) is  $P_{\rm RF-refr, 2K} = 305\times DC$  MW, it does not depend on $N$.
        \item  Electric power due to HOM losses, it is $P_{\rm HOM, w.p.} \approx 240(N/10^9)^2 \times DC$,  MW.
        \ei
The total power (only main contributions)
    \be
    P_{tot}= \left(20.8\frac{N}{10^{9}}+ 305+240 \left( \frac{N}{10^{9}} \right)^2\right) \times DC.
    \label{power}
    \ee
This number are for $d=23$ cm. Most of power is spent on removal of RF and HOM losses.

Let us  first discuss the dependence of the maximum luminosity of the bunch distance $d$ and find the optimum $N$ and $DC$.

Neglecting power losses in wigglers, the power in operation with a duty cycle $DC$
\be
P= \left(a + b \frac{N^2}{d}\right)  DC,
\label{abd}
\ee
where coefficients  $a$ and $b$ describe RF and HOM losses, respectively, they are both proportional to the collider length (or $E_0$). The luminosity
\be
L \propto \frac{N}{d} DC =\frac{N}{d}\left( \frac{P}{a+ bN^2/d}\right).
\ee
The maximum luminosity $L$
\be
L\propto \frac{P}{\sqrt{abd}}\;\;\;\; {\rm at} \;\;\;\; N=\sqrt{\frac{ad}{b}}, \;\;\;\; DC=\frac{P}{2a}.
\label{lumdc}
\ee
The luminosity reaches the maximum when the energy spent for removal of $RF$ and $HOM$ losses are equal (that is only for $DC<1$).
We see that $L\propto 1/\sqrt{d}$, so the distance between bunches $d$ should be as small as possible, that is why we assumed this in all above considerations. Second, the coefficient $a \propto 1/(\epsilon Q_0 T)$, where T is the temperature of SC cavities, $\epsilon$ is the technical efficiency (0.18 at T=1.8 K, 0.3 at T > 4 K), therefore $L \propto \sqrt{\epsilon Q_0 T}$. The optimum number of particles in the bunch does not depends on P and beam energy (because both $a$ and $b$ are proportional to $E$).

 According to the above power estimates for $2E_0=250$ GeV,  $a=305(E_0/125)$ MW, $b/d=240(E_0/125)/(10^9)^2$ MW, therefore the optimum number of particles in the bunch
 \be
 N/10^9 \approx \sqrt{305/240}=1.13.
 \ee
     Let's consider now the same for the CW operation. Now $DC=1$ and
 \be
 P= (a + b \frac{N^2}{d}), \;\;\; L \propto \frac{N}{d},
 \ee
  that gives
\be
N=\sqrt{\frac{(P-a)d}{b}}, \;\;\;\;\; L\propto \sqrt{\frac{(P-a)}{bd}}.
\label{lumcw}
\ee
Again $L\propto 1/\sqrt{d}$. The minimum power for CW operation $P=a \propto 1/(\epsilon Q_0 T)$. The number of particle in the bunch in the CW mode depends on the available power:
\be
 N/10^9 \approx \sqrt{\frac{(P-a)d}{b}} \sim \sqrt{\frac{(P(125/E_0) -305}{240}}
\ee
The luminosity in the CW mode is proportional to $\sqrt{P-a}$, at $P=2a$ it becomes equal to the maximum luminosity with a duty cycle ($DC=1$ in (\ref{lumdc})). It has sense to work at $P$ exceeding the threshold power only by about 35\% when $L_{\rm CW}/L_{\rm DC, max}=0.85$. In this case  $N/10^9 \sim 0.67$ and the power required for the Higgs factory is $P=410$ MW, it's too much.

Now all parameters are known, the resulting luminosities, calculated numerically with account of the neglected SR term are shown in the Fig.~\ref{luminosity}. For the Higgs factory with $2E=250$ GeV there are, for example, two options:
\bi
\item a duty cycle mode, $P=120$ MW, $L=0.39\times 10^{36}\,\cms$, $DC \approx 0.19$;
\item a continuous  mode, $P=410$ MW, $L=1.13\times 10^{36}\,\cms$.
\ei
The continuous mode is very attractive, but the required power is too high. Below we will discuss how it can be reduced.

The above numbers are for $Q_0=3\times 10^{10}$ and $T=1.8$ K. If $T=4.5$ K (Nb$_3$Sn or other) and $Q_0$ is the same, then $\epsilon T$ is 4 times larger, one can have  $L=0.6\times 10^{36}$ in CW mode already at $P=100$ MW, instead of 410 MW, that would be great.
\section{Ways to reduce power consumption}
The main power is spent on heat removal from RF and HOM losses. Two ways are seen: a) the use of a superconductor with a higher operating temperature, and) a decrease in the RF frequency \underline{$f \equiv f_{\rm RF}$ (in this section)}. A promising material is Nb$_3$Sn, which operates at a temperature of 4.5 K, where the efficiency of heat removal is about 4 times higher than for Nb at T = 1.8 K (apart from the Carnot efficiency, the technical efficiency is also about 1.6 times higher)~\cite{Posen2017}. Its thermal conductivity is about 1000 times lower than that of niobium, so it is used in the form of a thin film on material with high thermal conductivity, such as niobium or copper. Cavities with Nb$_3$Sn reach the same high $Q_0$ values as with Nb, although the technology is not yet reliable enough. As for Nb, the value of $Q_0 \propto 1/R_s$ is limited by the BCS surface conductivity $R_{\rm BCS}\propto f^2$; therefore, it is advantageous to lower the RF frequency.

So, the transition from T = 1.8 K to T = 4.5 K increases the efficiency of heat removal by a factor of 4. Let us denote this factor by the letters \eT. The RF power loss in cavities per unit length $P_{\rm RF} \propto R_Sf^{-1} \propto f$ (if $R_s=R_{\rm BCS}$). In addition, a decrease of $f$ leads to a decrease of HOM losses (per unit length): $P_{\rm HOM} \propto 1/r_a^2 \propto f^2$. The minimum distance between bunches $d \propto 1/f$. As a result, the parameters in the Eq.~\ref{abd} have the following dependence:
$a \propto f/(\eT),\, b\propto f^2,\, d \propto 1/f$. The luminosity for operation with DC, Eq.~\ref{lumdc},
\be
 L \propto \frac{P}{\sqrt{abd}} \propto P\frac{\sqrt{\eT}}{f}, \;\;\;\;\;\;\frac{L}{P}\propto\frac{\sqrt{\eT}}{f}.
\ee
With $\eT = 4$ and a 2-fold decrease of $f$, we obtain a 4-fold increase of the luminosity at the same power.

For continuous operation (Eq.~\ref{lumcw}) $L\propto \sqrt{(P-a)/(bd)}$, the threshold power $P\propto a \propto f/\eT$. At $\eT=4$ and 2-fold decrease of $f$ it decreases 8 times! Earlier we obtained the required power for CW operation at 250 GeV equal to 410 MW, it will decrease to 50 MW, which is already acceptable.

In CW mode $L \propto \sqrt{a/bd} \propto 1/\sqrt{\eT}$ at $P \sim a \propto f/\eT$, that gives $L/P= \sqrt{\eT}/f$,  the same gain as when working with DC.

 Thus, a transition from $T=1.8$ K to $T = 4.5$ K and a halving of the RF-frequency increases the luminosity by a factor of 4 and decreases the threshold power for continuous operation by a factor of 8. This is in the ideal case when the surface conductivity  $\sigma_s \approx \sigma_{\rm BSC}$. Such collider variants are also presented in the summary table.
\section{From 250 to 500 GeV} \vspace{-2mm}
     In this article, the main focus was on the high luminosity 250 GeV Higgs factory.   An increase in luminosity gives a sensitivity to masses of new particles responsible for deviations of the Higgs coupling constants from the SM.
      However, there is also a great interest to higher energies: $2E_0\!=\!360$ GeV (the top-quark threshold), $2E_0\!=\!500$ GeV (the Higgs self-coupling).

        Beside collision effects there is also a problem of emittance dilution in the beam delivery where horizontal dispersion is required for chromatic correction at the final focus. In the ERLC, the beam pass this region about 400 times during the damping time. Some increase of the length will be needed to solve this problem.

        The continuous operation required twice more threshold power than at $2E_0\!=\!250$ GeV. The CW mode will be realistic in case of success with Nb$_3$Sn cavities.

        The expected luminosities at $2E_0\!=\!500$ GeV are given in the Table~\ref{Table3} and Fig.~\ref{luminosity}, they are about 3 times lower than at $2E_0\!=\!250$ GeV for the same total powers.
\vspace{-2mm}
\section{Summary tables}
\begin{figure}[pt!]
\centering
\includegraphics[width=13.cm,height=11.cm]{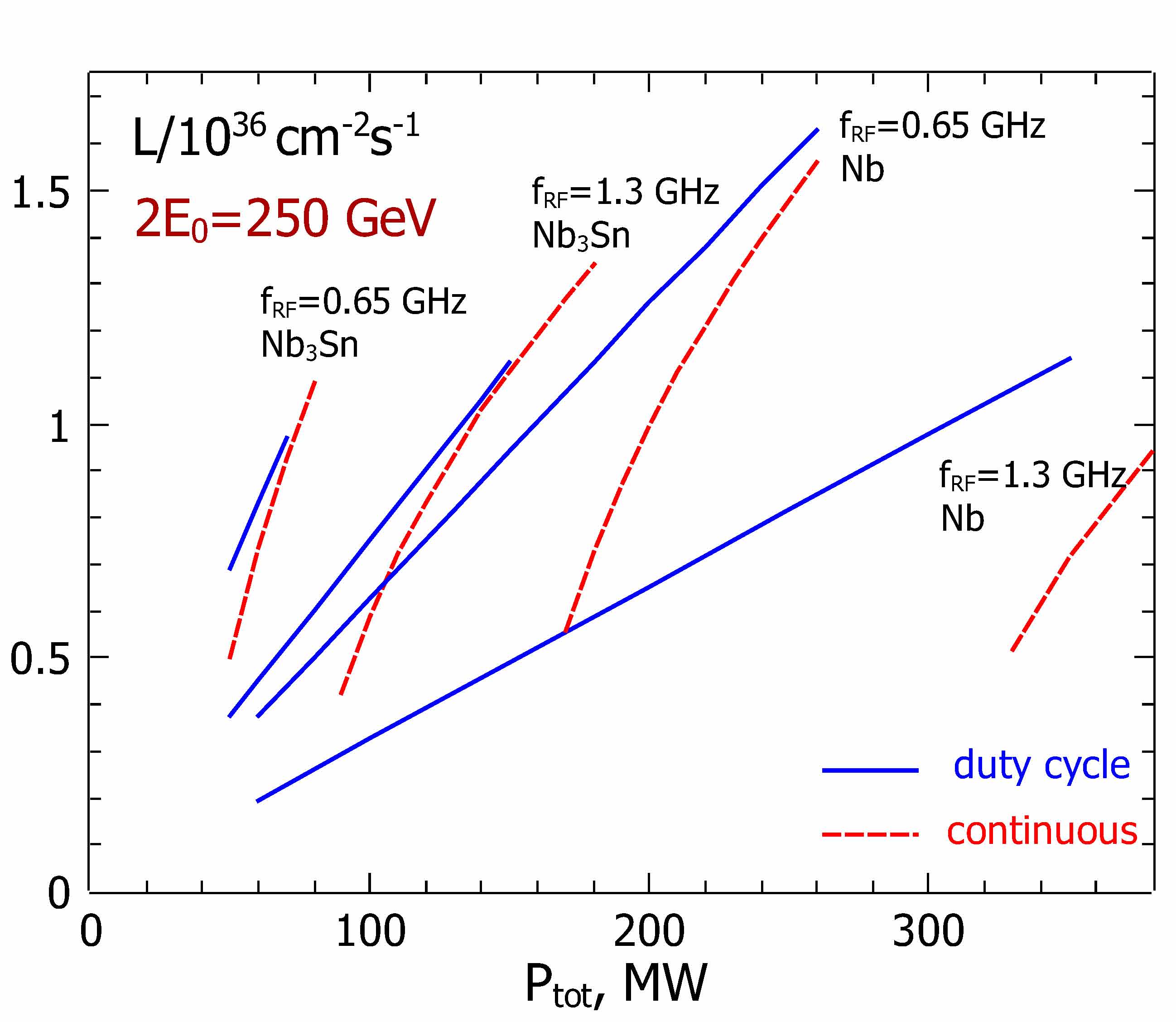}
\includegraphics[width=13.cm,height=11.cm]{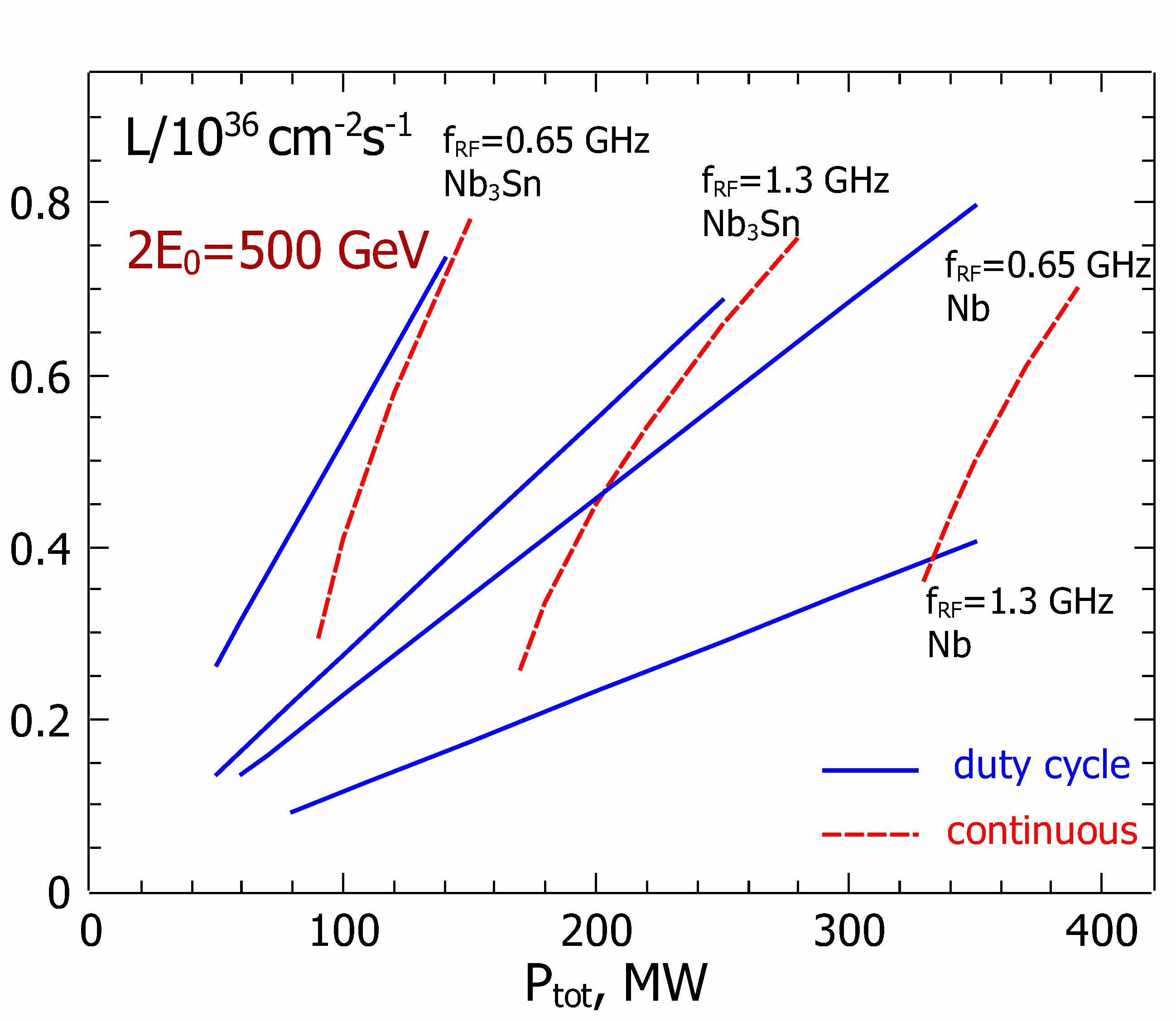}
\caption{Dependence of the luminosity on the total power for $2E_0=250$ GeV (upper) and $2E_0=500$ GeV (bottom); blue (solid) line ---  optimum duty cycle operation, red (dashed) curves --- continuous (CW) operation, see the text.}
\label{luminosity}
\end{figure}

\vspace{-2mm}
  In this article on the new linear collider scheme, only the most important problems affecting luminosity are considered,  there are many  issues that require careful consideration by accelerator experts.  The preliminary parameters of the collider with the energy $2E_0 =$ 250 and 500 GeV  are presented in Tables \ref{Table2} and \ref{Table3}. Each table contains 4 ERLC options and ILC.\footnote{Please note, for the ILC the total power is given in the tables. Linac itself consumes about 1/3 at $2E_0=250$ GeV (large fraction add sources and damping rings) and 2/3 at $2E_0=500$ GeV. }  The dependence of the luminosities on the total power for various options is shown in the Fig.~\ref{luminosity}.
  Beam emittances are chosen similar to the ILC (just because 5 GeV arcs with wigglers looks like the ILC DR).  The ERLC and ILC consist of the same elements: linear accelerators, arcs, compressors, but in the ILC the bunch passes this way once, while in the ERLC  the damping time corresponds to about 400 revolution. Bunch compressors are of greatest concern, a decrease of the beam energy in the return loop may be required to reduce the emittance dilution.  There are questions on transverse beam dynamics due to the HOMs. Hope the problem is not serious because the bunch charge is about 20 times lower than at the ILC. The contributions of the main energy consumers for two ERLC options at $2E_0=250$ GeV are shown in Table~\ref{Table4}.

\begin{table}[!hbtp]
\caption{Parameters of \epem linear colliders ERLC and ILC, $2E_0=250$ GeV.}
{\renewcommand{\arraystretch}{0.8} \setlength{\tabcolsep}{1mm}
\begin{center} 
\begin{tabular}{ l  l  c  c  c  c  c }  \hline \\[-0.3cm]
                        & unit           &                      ERLC  & ERLC     &ERLC     & ERLC    &ILC \\ [-1.1mm]
                        &                  &                   pulsed & pulsed   & contin. & contin.  &     \\
                        &                  &                   Nb     & Nb      & $\rm Nb_3Sn$ & $\rm Nb_3Sn$ & Nb   \\
                        &                  &                   1.8 K  & 1.8 K   & 4.5 K    &  4.5 K  & 1.8 K    \\
                        &                  &                   1.3 GHz& 0.65 GHz& 1.3 GHz  &  0.65 GHz& 1.3 GHz \\  \hline \\[-2mm]
Energy $2E_0$           &GeV             &                      250   &  250     & 250     &250      &250      \\
Luminosity ${\mathcal L}_{{\rm tot}}$ &$10^{36}$\cms &         0.39   &  0.75    & 0.83    & 1.6     & 0.0135\\
$P$ (wall) (collider)   &MW              &                    120     &120       & 120     & 120     & 129(tot.)\\
Duty cycle, $DC$        &                &                     0.19  &  0.37    &   1     &  1      & n/a\\
Accel. gradient, $G$    &MV/m            &          20                & 20       &20       &20       & 31.5 \\
Cavity quality, $Q$     & $ 10^{10}$     &                    3       &  12      & 3       & 12      & 1     \\
Length $L_{\rm act}/L_{\rm tot}$ &km     &                     12.5/30&12.5/30   & 12.5/30 & 12.5/30 & 8/20 \\
$N$ per bunch           &$10^{9}$       &                     1.13    & 2.26     & 0.46    & 1.77    & 20\\
Bunch distance          & m              &                   0.23     &0.46      & 0.23    & 0.46   & 166 \\
Rep. rate, $f$          & Hz             &          $2.47\cdot 10^8$ &$2.37\cdot 10^8$&$1.3 \cdot 10^9$  &$6.5\cdot 10^8$ & 6560 \\
$\epsilon_{x,\,n}$/$\epsilon_{y,\,n}$ & $10^{-6}$ m &        10/0.035 &10/0.035   & 10/0.035 &10/0.035  & 5/0.035 \\
$\beta^*_x$/$\beta_y$ at IP           & cm    &        2.7/0.031     & 10.8/0.031 &0.46/0.031&6.8/0.031 & 1.3/0.04 \\
$\sigma_x$ at IP       &  $\mum$        &                     1.05    &2.1      &0.43     &1.66     & 0.52\\
$\sigma_y$ at IP       & nm             &                   6.2       & 6.2      &6.2      &6.2      & 7.7\\
$\sigma_z$ at IP       & cm             &                   0.03      &0.03      &0.03     &0.03     & 0.03\\
$ (\sigma_E/E_0)_{\rm BS}$ at IP  & $\%$        &             0.2      &0.2       &0.2      &0.2      & $\sim 1$ \\ \hline
\end{tabular}
\end{center}
\vspace{-0mm}
}
\label{Table2}
\end{table}
\begin{table}[!hbtp]
\vspace{-1cm}
\caption{Parameters of \epem linear colliders ERLC and ILC, $2E_0=500$ GeV.}
{\renewcommand{\arraystretch}{0.8} \setlength{\tabcolsep}{1mm}
\begin{center} 
\begin{tabular}{ l  l  c  c  c  c  c }  \hline \\[-0.3cm]
                        & unit           &                      ERLC  & ERLC     &ERLC     & ERLC    &ILC \\ [-1.1mm]
                        &                  &                   pulsed & pulsed   & pulsed  & contin.  &     \\
                        &                  &                   Nb     & Nb      & $\rm Nb_3Sn$ & $\rm Nb_3Sn$ & Nb   \\
                        &                  &                   1.8 K  & 1.8 K   & 4.5 K    &  4.5 K  & 1.8 K    \\
                        &                  &                   1.3 GHz& 0.65 GHz& 1.3 GHz  &  0.65 GHz& 1.3 GHz \\  \hline \\[-2mm]
Energy $2E_0$           &GeV             &                      500   &  500     & 500     & 500      &500      \\
Luminosity ${\mathcal L}_{{\rm tot}}$ &$10^{36}$\cms &         0.174   &  0.342    & 0.412    & 0.78    & 0.018\\
$P$ (wall) (collider)   &MW              &                    150     &150       & 150     & 150     & 163(tot) \\
Duty cycle, $DC$        &                &                     0.121  &  0.237   &  0.47   &  1      & n/a\\
Accel. gradient, $G$    &MV/m            &          20                & 20       &20       &20       & 31.5 \\
Cavity quality, $Q$     & $ 10^{10}$     &                    3       &  12      & 3       & 12      & 1     \\
Length $L_{\rm act}/L_{\rm tot}$ &km     &                     25/50  & 25/50    & 25/50   & 25/50   & 16/31 \\
$N$ per bunch           &$10^{9}$       &                     1.13    & 2.26     & 0.685    & 1.23    & 20\\
Bunch distance          & m              &                   0.23     &0.46      & 0.23     &  0.46   & 166 \\
Rep. rate, $f$          & Hz             &           $1.57\cdot 10^8$ &$1.54\cdot 10^8$&$6.1 \cdot 10^8$ &$6.5\cdot 10^8$ & 6560 \\
$\epsilon_{x,\,n}$/$\epsilon_{y,\,n}$    &   $10^{-6}$ m &  10/0.035  &10/0.035  & 10/0.035 &10/0.035  & 10/0.035 \\
$\beta^*_x$/$\beta_y$  at IP      & cm               &     7.7/0.089 & 31/0.089 &2.85/0.089 &9.4/0.089 & 1.1/0.04 \\
$\sigma_x$ at IP       &  $\mum$        &                     1.26    &2.5      &0.76     &1.38     & 0.47\\
$\sigma_y$ at IP       & nm             &                   7.4       & 7.4      &7.4      &7.4      & 5.9\\
$\sigma_z$ at IP       & cm             &                   0.089      &0.089    &0.089    &0.089    & 0.03\\
$ (\sigma_E/E_0)_{\rm BS}$ at IP  & $\%$        &             0.1      &0.1       &0.1      &0.1      & $\sim 1$ \\ \hline
\end{tabular}
\end{center}
\vspace{-0mm}
}
\label{Table3}
\end{table}

\begin{table}[!hbtp]
\caption{Power consumption of two ERLC options at $2E_0=250$ GeV, presented in the first and fourth columns of the Table~\ref{Table2}.}
{\renewcommand{\arraystretch}{1.2} \setlength{\tabcolsep}{3 mm}
\begin{center}
\begin{tabular}{ l   c  c }  \hline
                        & Nb, 1.3 GHz, T=1.8 K                    & Nb$_3$Sn, 0.65 GHz, T=4.5 K         \\
                        & $L=0.39\times 10^{36}$ \cms             & $L=1.6\times 10^{36}$ \cms           \\
                        & $N=1.13\times10^9$, $DC=0.19$           & $N=1.77\times10^9$, $DC=1$   \\ \hline
Beam generation         & small  & small  \\
Radiation in wigglers   & 4.45                                      & 18.4  \\
HOMs, beam energy       & 5.5                                     & 9  \\
HOMs cool., 1.8(4.5) K     & 24.8                                    & 10  \\
HOMs cool., 77 K      & 27.6                                    & 44.7   \\
RF diss.cool., 1.8(4.5) K   & 57.6                                    & 38  \\ \hline
$P_{tot}$, MW           & 120                                     & 120  \\ \hline
\end{tabular}
\end{center}
\vspace{-0mm}
}
\label{Table4}
\end{table}

\vspace{-2mm}
\section{Conclusion} \vspace{-2mm}
   Currently, the design of superconducting ILC is very similar to that of any room-temperature LC: the  beams are used only once and superconductivity does not add much (a slight increase in efficiency, larger distances between bunches, looser tolerances and a lower peak klystron power). The luminosities are about the same.  This scheme was laid down 40 years ago. Even earlier, there were proposals to use energy recovery in SC linear colliders, but they were considered unattractive, since the expected luminosity was much lower than in single-pass colliders.

  In this article, I propose a way to overcome the main obstacle faced by SC linear colliders with  energy recovery: parasitic collisions in linacs. The proposed  scheme of a {\it twin} linear collider opens the way to an energy-recovery LC with multiple use of beams. At $2E_0= $250--500 GeV, the possible luminosity is up to two order of magnitude higher than at the ILC and much higher than at the FCC. To achieve the best performance, additional R\&D efforts are required to develope superconducting  cavities that reliably operate at 4.5 K with high $Q_0$ values.
\section*{Acknowledgment}
 This work was supported by the Russian Foundation for Basic Research grant RFBR 20-52-12056. I would like to thank U.~Amaldi, A.~Bondar, A.~Hutton, N.~Dikansky, M.~Klein, E.~Levichev, P.~Logatchev, I.~Meshkov, V.~Parkhomchuk, F.~Richard, B.~Sharkov, A.~Skrinsky, S.~Stapnes and N.~Vinokurov  for their interest to this proposal and Akira Yamamoto and Kaoru Yokoya for useful remarks. Following my talk at the LCWS-2021, a dedicated HEP sub-panel was formed to evaluate this proposal. It is very important to quickly understand the feasibility of the project and, if everything is realistic, move towards the goal.

\end{document}